\documentstyle[12pt,epsfig]{article}
\begin{document}
\begin{center}
\vspace{1cm}{\large\bf Effect of field quantization on Rabi
oscillation of equidistant cascade four-level system}\\
\vspace{.9in}{\bf Mihir Ranjan Nath{\footnote[1]
{mrnath\_95@rediffmail.com}}, Tushar Kanti Dey, {\footnote[2]
{tkdey54@rediffmail.com}} Surajit Sen{\footnote[3]
{ssen55@yahoo.com}}\\Department of Physics\\
Guru Charan College\\ Silchar 788004, India}
\begin{center}
and
\end{center}
\vspace{.5cm} {\bf Gautam Gangopadhyay
{\footnote[4]{gautam@bose.res.in}}\\
S N Bose National Centre for Basic Sciences\\
JD Block, Sector III
\\Salt Lake City, Kolkata 700098, India.}\\
\end{center}
\vspace{1cm}
\begin{abstract}
We have exactly solved  a model of equidistant cascade four-level
system interacting  with a single-mode radiation field both
semiclassically and quantum mechanically by exploiting its
similarity with Jaynes-Cummings model. For the classical field, it
is shown that the Rabi oscillation of the system initially in the
first level (second level) is similar to that of the system when it
is initially in the fourth level (third level). We then proceed to
solve the quantized version of the model where the dressed state is
constructed by using a six parameter four-dimensional matrix and
show that the symmetry exhibited in the Rabi oscillation of the
system for the semiclassical model is completely destroyed on the
quantization of the cavity field. Finally we have studied the
collapse and revival of the system for the  cavity field-mode in a
coherent state to discuss the restoration of symmetry and its
implication is discussed. \vspace{.5in} \vfill

\end{abstract}

\vfill
 \pagebreak

\begin{center}
\large I.Introduction
\end{center}
\par
Over the decades, the theory of Electron Spin Resonance (ESR) has
been regarded as the key model to understand various fundamental
aspects of the semiclassical two-level system [1]. Its fully
quantized version, namely, the two-level Jaynes-Cummings model (JCM)
has also been proven to be an useful theoretical laboratory to address
many subtle issues of the light-atom interaction which eventually
gives birth to the cavity electrodynamics [1,2]. A natural but
non-trivial extension of the JCM is the three-level system and it
exhibits wide verity of quantum-optical phenomena such as two-photon
coherence [3], resonance Raman scattering [4], double resonance
process [5], population trapping [6], three-level super radiance
[7], three-level echoes [8], STIRAP [9], quantum jump [10], quantum
zeno effect [11] etc. As a straight forward generalization of the
three-level system, the multi-level system interacting with
monochromatic laser is also extensively investigated [12-17]. Thus
it is understood that the increase of the number of level leads to
the emergence of a plethora of phenomena and the upsurge
of ongoing investigations 
of the four-level system is undoubtedly to predict more phenomena.
For example, out of different configurations of the four-level
system, the tripod configuration has come into the purview of recent
studies particularly because it exhibits the phenomenon of the
Electromagnetically Induced Transparency (EIT) [18-25] which also
received experimental confirmation [26-29]. Such system is proposed
to generate the non-abelian phases [30], qubit rotation [31],
coherent quantum switching [32], coherent controlling of nonlinear
optical properties [33], embedding two qubits [34] etc. These
developments lead to the careful scrutiny of all possible
configurations of the four-level system including the cascade
four-level system which we shall discuss here.
\par
In recent past the equidistant cascade four-level system interacting
with the semiclassical and quantized field was discussed mainly
within the frame work of generalized N-level system [35-40]. Other
variant of this configuration, often referred to as Tavis-Cummings
model, is studied to construct possible controlled unitary gates
relevant for the quantum computation [41,42]. However, these
treatments are not only devoid of the explicit calculation of the
probabilities for all possible initial conditions [43,44], but they
also bypass the comparison between the semiclassical and the
quantized models which is crucial to discern the exact role of the
field quantization on the population oscillation. In this work we
have developed a dressed atom approach of calculating the
probabilities with all possible initial conditions especially in the
spirit of the basic theory of the ESR model and JCM taking the field
to be either monochromatic classical or quantized field [1]. This
work is the natural extension of our previous works on the
equidistant cascade three-level model [44], where it is explicitly
shown that the symmetric pattern observed in the population dynamics
for the classical field is completely spoilt on the quantization of
the cavity mode.
\par
The remaining sections of the paper are organized as follows. In
section-II we discuss the equidistant cascade four-level system
modeled by the generators of the spin-$\frac{3}{2}$ representation
of $SU(2)$ group and then study its Rabi oscillation with different
initial conditions taking interacting field to be the classical
field. Section-III deals with the solution of the four-level system
taking the cavity field mode to be the quantized mode. In Section-IV
we compare the Rabi oscillation of the system of the semiclassical
model with that of the quantized field and discuss the collapse and
revival phenomenon and its implications. Finally in Section-V we
summarize our results and discuss the outlook of our investigation.
\pagebreak
\begin{center}
\large II. The Semiclassical Cascade Four-Level System
\end{center}
\par
The Hamiltonian of the equidistant cascade four-level system is
given by
\begin{equation}
 H(t)=\hbar\omega_0 J_3 + \hbar \kappa (J_ + \exp ( -
i\Omega  t) + J_ - \exp (i\Omega  t)),
\end{equation}
where $J_+$, $J_-$ and $J_3$ be the generators of the
spin-$\frac{3}{2}$ representation of $SU(2)$ group given by
\begin{equation}
     J_ +   = \left[ {\begin{array}{*{20}c}
   0 & \sqrt{3} & 0 & 0  \\
   0 & 0 & 2 & 0  \\
   0 & 0 & 0 & \sqrt{3}  \\
   0 & 0 & 0 & 0  \\
\end{array}} \right],  \\
    J_-   = \left[ {\begin{array}{*{20}c}
   0 & 0 & 0 & 0  \\
  \sqrt{3} & 0 & 0 & 0  \\
   0 & 2 & 0 & 0  \\
   0 & 0 & \sqrt{3} & 0  \\
\end{array}} \right],  \\
    J_3 =   \left[{\begin{array}{*{20}c}
  \frac{3}{2} & 0 & 0 & 0 \\
   0 & \frac{1}{2} & 0 & 0 \\
   0 & 0 & -\frac{1}{2} & 0 \\
   0 & 0 & 0 & -\frac{3}{2} \\
\end{array}} \right]
 \end{equation}
In Eq.(1), ${\hbar}\omega_0$ be the equidistant energy gap between
the levels, $\Omega$ be the frequency of the classical mode and
$\kappa$ be the coupling constant of the light-atom interaction
respectively. The time evolution of the system is described by the
Schr\"odinger equation
\begin{equation}
 i\hbar \frac{{\partial \Psi }}{{\partial
t}} = { H(t)}\Psi,
\end{equation}
where the above time-dependent Hamiltonian in the matrix form is
given by

\begin{equation}
 H(t)   = \left[ {\begin{array}{*{20}c}
   \frac{3}{2}\hbar\omega_0 & \sqrt{3}\hbar\kappa \exp[-i\Omega t] & 0
 & 0    \\
   \sqrt{3}\hbar \kappa \exp[i\Omega t] & \frac{1}{2}\hbar\omega_0 &
   2 \hbar\kappa \exp[-i\Omega t] & 0  \\
   0 & 2 \hbar\kappa \exp[i\Omega t] & -\frac{1}{2}\hbar\omega_0 &
   \sqrt{3}\hbar\kappa \exp[-i\Omega t]  \\
   0 & 0 & \sqrt{3}\hbar\kappa \exp[i\Omega t] &
   -\frac{3}{2}\hbar\omega_0 \\
\end{array}} \right].
\end{equation}
To find the amplitudes, let the solution of the Schr\"{o}dinger
equation corresponding to this Hamiltonian is given by
\begin{equation}
 \Psi (t) =
C_1(t) \left| 1 \right\rangle + C_2(t) \left| 2 \right\rangle  +
C_3(t) \left| 3 \right\rangle + C_4(t) \left| 4
\right\rangle,\end{equation} where $C_1(t)$, $C_2(t)$ , $C_3(t)$ and
$C_4(t)$ are the time-dependent normalized amplitudes with basis
states

\begin{equation}
\left| 1 \right\rangle  = \left[ {\begin{array}{*{20}c}
   0  \\
   0  \\
   0  \\
   1  \\
\end{array}} \right],
\left| 2 \right\rangle  = \left[ {\begin{array}{*{20}c}
   0  \\
   0  \\
   1  \\
   0  \\
\end{array}} \right],
\left| 3 \right\rangle  = \left[ {\begin{array}{*{20}c}
   0  \\
   1  \\
   0  \\
   0  \\
\end{array}} \right],
\left| 4 \right\rangle  = \left[ {\begin{array}{*{20}c}
   1  \\
   0  \\
   0  \\
   0  \\
\end{array}} \right],
\end{equation}
respectively. The Schr\"odinger equation in Eq.(3) can be written as
\begin{equation}
 i\hbar \frac{{\partial \tilde \Psi }}{{\partial t}}=\tilde
{H}\tilde \Psi,
\end{equation}
where the time-independent Hamiltonian is given by
\begin{equation}
\widetilde{H} =  - i\hbar U^ \dagger  \dot U + U^
\dagger { H(t)}U,
\end{equation}
with the unitary operator $U(t) = e^{-i{\Omega}J_3 t}$. The rotated
wave function appearing in Eq.(7) is obtained by the unitary
transformation {\setlength\arraycolsep{2pt}
\begin{eqnarray}
\nonumber
\widetilde{\Psi}(t) &=& U(t)^{\dag}\Psi(t) \\
&=& e^{-i\frac{3}{2}{\Omega} t} C_1(t) \left| 1 \right\rangle +
e^{-i\frac{1}{2}{\Omega}t} C_2(t) \left| 2 \right\rangle +
e^{i\frac{1}{2}{\Omega} t} C_3(t) \left| 3 \right\rangle +
e^{i\frac{3}{2}{\Omega} t} C_4(t) \left| 4 \right\rangle.
\end{eqnarray}}
We thus note that the amplitudes are simply modified by a phase term
and hence do not contribute to the probabilities. The
time-independent Hamiltonian in Eq.(8) is given by
\begin{equation}
\widetilde{H} = \hbar\left [ {\begin{array}{*{20}c}
   {\frac{3}{2} \Delta } & {\sqrt{3}\kappa  } & 0 & 0  \\
   {\sqrt{3}\kappa  } & {\frac{1}{2}\Delta } & {2\kappa  } & 0  \\
   0 & {2 \kappa  } & { - \frac{1}{2}\Delta } & {\sqrt{3}\kappa  }  \\
   0 & 0 & {\sqrt{3}\kappa  } & { - \frac{3}{2}\Delta }  \\
\end{array}} \right],
\end{equation}
where $\Delta=\omega_0-\Omega$. At resonance $(\Delta=0)$, the eigen
values of the Hamiltonian are given by
$\lambda_{1}=-\lambda_{4}=-3\hbar\kappa$ and
$\lambda_{2}=-\lambda_{3}=-\hbar\kappa$ respectively which can also
be generated by the transformation
\begin{equation}
diag(\lambda_1, \lambda_2, \lambda_3, \lambda_4) =
T_\alpha\widetilde{H} T_\alpha^{-1},
\end{equation}
where $T_\alpha$ be the transformation matrix given by
\begin{equation}
 {T_\alpha} = \left[ {\begin{array}{*{20}c}
   {\alpha _{11} } & {\alpha _{12} } & {\alpha _{13} } & {\alpha _{14}
 }  \\
   {\alpha _{21} } & {\alpha _{22} } & {\alpha _{23} } & {\alpha _{24}
 }  \\
   {\alpha _{31} } & {\alpha _{32} } & {\alpha _{33} } & {\alpha _{34}
 }  \\
   {\alpha _{41} } & {\alpha _{42} } & {\alpha _{43} } & {\alpha_{44} }
  \\
\end{array}} \right].
\end{equation}
The different elements of  matrix, which preserves the
orthogonality, are given by [46] {\setlength\arraycolsep{2pt}
\begin{eqnarray}
\nonumber \alpha _{11}&=& c_1c_5+s_1s_3s_4s_5\\
\nonumber \alpha _{12}&=& -c_1s_5s_6+s_1c_3c_6+s_1s_3s_4c_5s_6\\
\nonumber \alpha _{13}&=& s_1s_3c_4\\
\nonumber \alpha _{14}&=&
-c_1s_5c_6-s_1c_3s_6+s_1s_3s_4c_5c_6\\
\nonumber \alpha _{21}&=& -s_1c_2c_5+(c_1c_2s_3-s_2c_3)s_4s_5\\
\nonumber \alpha _{22}&=&
s_1c_2s_5s_6+(c_1c_2c_3+s_2s_3)c_6+(c_1c_2s_3-s_2c_3)s_4c_5s_6\\
\nonumber \alpha_{23}&=&(c_1c_2s_3-s_2c_3)c_4\\
\nonumber \alpha _{24} &=&
s_1c_2s_5c_6-(c_1c_2c_3+s_2s_3)s_6+(c_1c_2s_3-s_2c_3)s_4c_5c_6\\
\nonumber \alpha _{31}&=&-s_1s_2c_5+(c_1s_2s_3+c_2c_3)s_4s_5\\
\nonumber \alpha_{32}&=& s_1s_2s_5s_6+(c_1s_2c_3-c_2s_3)c_6+(c_1s_2s_3+c_2c_3)s_4c_5s_6\\
\nonumber \alpha _{33}&=& (c_1s_2s_3+c_2c_3)c_4\\
\nonumber \alpha _{34}&=&
s_1s_2s_5c_6-(c_1s_2c_3-c_2s_3)s_6+(c_1s_2s_3+c_2c_3)s_4c_5c_6\\
\nonumber \alpha _{41}&=& c_4s_5\\
\nonumber \alpha _{42}&=& c_4c_5s_6\\
\nonumber \alpha _{43}&=& -s_4\\
\alpha _{44}&=& c_4c_5c_6
\end{eqnarray} }
where $s_i=\sin\theta_i$ and  $c_i=\cos\theta_i$ ($i=1,2,3,4,5,6$).
A straightforward calculation gives various angles to 
 matrix to be
{\setlength\arraycolsep{2pt}
\begin{eqnarray}
\begin{array}{*{20}l}
 \theta_1 &=& \arccos(-\sqrt{\frac{2}{5}}), &\theta_2 &=& \frac{3\pi}{4},
 & \theta_3 &=& -\frac{\pi}{2},\\
 \theta_4 &=& -\arcsin(\sqrt{\frac{3}{8}}), &\theta_5 &=& \arcsin
(\sqrt{\frac{1}{5}}), &\theta_6 &=& \frac{\pi}{3}.
\end{array}
\end{eqnarray}}
The time-dependent probability amplitudes of the four-levels are
given by
\begin{eqnarray}
 \left[ {\begin{array}{*{20}c}
   {C_1 (t)}  \\
   {C_2 (t)}  \\
   {C_3 (t)}  \\
   {C_4 (t)}  \\
\end{array}} \right] = T_\alpha^{ - 1} \left[ {\begin{array}{*{20}c}
   {e^{ - i\lambda _1 t} } & 0 & 0 & 0  \\
   0 & {e^{ - i\lambda _2 t} } & 0 & 0  \\
   0 & 0 & {e^{ - i\lambda _3 t} } & 0  \\
   0 & 0 & 0 & {e^{ - i\lambda _4 t} }  \\
\end{array}} \right]T_\alpha\left[ {\begin{array}{*{20}c}
   {C_1 (0)}  \\
   {C_2 (0)}  \\
   {C_3 (0)}  \\
   {C_4 (0)}  \\
\end{array}} \right].
\end{eqnarray}

Later in section IV, we proceed to analyze the probabilities of the
four levels numerically for four distinct initial conditions,
namely,
\begin{eqnarray}
\nonumber
 {\begin{array}{*{20}l}
\textrm{Case - I} &: & C_1 (0) = 1, & C_2 (0) = 0, & C_3 (0) = 0, & C_4 (0) = 0,\\
\textrm{Case - II} &: & C_1 (0) = 0, &C_2 (0) = 1, & C_3 (0) = 0, & C_4 (0) = 0,\\
\textrm{Case - III} &: & C_1 (0) = 0, & C_2 (0) = 0, & C_3 (0) = 1, & C_4 (0) = 0,\\
\textrm{Case - IV} &: & C_1 (0) = 0, &C_2 (0) = 0, & C_3 (0) = 0, &
C_4 (0) = 1, \end{array}}
\end{eqnarray}
respectively.

\begin{center}
\large III. The Jaynes-Cummings Model of Cascade Four-Level System
\end{center}
\par
We now consider the equidistant cascade four-level system
interacting with a mono-chromatic quantized cavity field. The
Hamiltonian of such system in the rotating wave approximation (RWA)
is given by

\begin{equation}
H= \hbar\Omega (  J_3 + { a^ \dagger } a) + \hbar(\Delta J_3 + g (J_
+ a + J_ - a^\dagger )).
\end{equation}
This is an archetype JCM where the Pauli matrices are replaced by
the spin-$\frac{3}{2}$ representation of $SU(2)$ group. Using the
algebra of the $SU(2)$ group and that of the field mode it is easy
to see that the two parts of the Hamiltonian shown in the
parenthesis of Eq.(16) commute with each other indicating that they
have the simultaneous wave function. Let the eigen function
 \begin{equation}
 \left| {\Psi_q(t)} \right\rangle  = \sum\limits_{n = 0}^\infty
  {[C_1^{n + 2} (t)\left| {n + 2,1} \right\rangle }  + C_2^{n + 1} (t)
  \left| {n + 1,2} \right\rangle  + C_3^n (t)\left| {n,3} \right\rangle
  + C_4^{n - 1} (t)\left| {n - 1,4} \right\rangle ],
 \end{equation}
where $n$ represents the number of photons in the cavity field. The
Hamiltonian couples the atom-field states ${\left| {n + 2,1}
\right\rangle }$, ${\left| {n + 1,2} \right\rangle }$, ${\left|
{n,3} \right\rangle }$, and ${\left| {n - 1,4} \right\rangle }$
respectively. At resonance $\Delta=0$, the interaction part of the
Hamiltonian in the matrix form is given by

\begin{equation}
 H_{\rm{int}} = g\hbar \left[ {\begin{array}{*{20}c}
   0 & {\sqrt {3(n + 2)} } & 0 & 0  \\
   {\sqrt {3(n + 2)} } & 0 & {2\sqrt {n + 1}} & 0  \\
   0 & {2\sqrt {n + 1} } & 0 & {\sqrt{3n} }  \\
   0 & 0 & {\sqrt {3n } } & 0  \\
\end{array}} \right],
\end{equation}
with the eigenvalues {\setlength\arraycolsep{2pt}
\begin{equation}
\begin{array}{*{20}l}
\lambda_{1q}&=&-\lambda_{4q}&=&-g \hbar \sqrt {5(1+n) + b},\\
\lambda_{2q}&=&-\lambda_{3q}&=&-g \hbar \sqrt {5(1+n) - b},
\end{array}
\end{equation}
respectively where $b=\sqrt{25+16n(2+n)}$. The dressed eigen states
are constructed by rotating the bare states as
\begin{equation}
 \left[ {\begin{array}{*{20}c}
   {\left| {n,1} \right\rangle }  \\
   {\left| {n,2} \right\rangle }  \\
   {\left| {n,3} \right\rangle }  \\
   {\left| {n,4} \right\rangle }  \\
\end{array}} \right] = T_n \left[ {\begin{array}{*{20}c}
   {\left| {n + 2,1} \right\rangle }  \\
   {\left| {n + 1,2} \right\rangle }  \\
   {\left| {n,3} \right\rangle }  \\
   {\left| {n - 1,4} \right\rangle }  \\
\end{array}} \right],
\end{equation}
where $T_n$ is similar to the aforementioned orthogonal
transformation matrix whose different elements are given by
{\setlength\arraycolsep{2pt}
\begin{eqnarray}
\begin{array}{*{20}l}
\alpha_{11} &= & -\alpha_{41}   &=&  -\frac{{(1+b-2n)\sqrt {5+ b+5n}
}}{{2\sqrt{3(2+n)\{5(5+b)+2n(16+b+8n)\} }}},\\  \alpha_{21}  &=&  -
\alpha{}_{31}  &=&
\frac{(b-1+2n)\sqrt{(5+5n-b)(5+2n+b)}}{12\sqrt{n(n+1)(n+2)b}}, \\
 \alpha_{12}  &= & \alpha_{42}  &=&
\frac{5+2n+b}{2\sqrt{5(5+b)+2n(16+b+8n)}}, \\ \alpha_{13}  &=&
-\alpha_{43} &=&
-\frac{\sqrt{(1+n)(5+b+5n)}}{\sqrt{5(5+b)+2n(16+b+8n)}},\\
\alpha_{22}  &=  & \alpha_{32}  &=&
-\frac{\sqrt{3n(1+n)}}{\sqrt{b(5+b+2n)}}
, \\ \alpha_{14}  &= & \alpha_{44} &=& \frac{\sqrt{(b-5-2n)}}{2\sqrt{b}},\\
\alpha_{23}  &= & -\alpha_{33}  &=&
-\frac{\sqrt{(5+5n-b)(5+2n+b)}}{2\sqrt{3nb}},  \\ \alpha_{24}  &=&
\alpha_{34}  &=& \frac{\sqrt{5+2n+b}}{2\sqrt{b}} .
\end{array}
\end{eqnarray}}

A straightforward but rigorous calculation gives the explicit
expressions of the angle of rotation for the quantized model

{\setlength\arraycolsep{2pt}
\begin{eqnarray}
\nonumber \theta_1 &=& \arccos\left[\frac{{-\alpha
_{11}}}{{\sqrt{(1-\alpha
_{13}^2)(1 - \alpha _{11}^2  - \alpha _{13}^2)} }}\right],\\
\nonumber \theta _2 &=& \arccos \left[\frac{{\alpha _{11} \alpha
_{13} \alpha _{23} + (1 - \alpha _{13}^2 )\sqrt {(1 - 2\alpha
_{11}^2  - 2\alpha _{13}^2 )(1 - 2\alpha _{13}^2  - \alpha _{23}^2
)} }}{{(2\alpha _{13}^2-1)\sqrt {(\alpha _{13}^2-1)^2 + \alpha
_{11}^2
(\alpha _{13}^2  - 2)} }}\right],\\
\nonumber  \theta _3  &=& \arcsin\left[\frac{{\alpha _{13} \sqrt
{\alpha _{11}^2 + \alpha _{13}^2  - 1} }}{{\sqrt {\alpha _{11}^2 (2
- \alpha
_{13}^2 )+(1 - \alpha _{13}^2 )^2 } }}\right],\\
\nonumber \theta_4 &=&\arcsin[\alpha _{13}],\\
\nonumber \theta _5  &=& -\arcsin \left[\frac{{\alpha _{11}
}}{{\sqrt {1 -
\alpha _{13}^2 } }}\right],\\
\theta_6 &=& \arcsin\left[\frac{{\alpha_{12}}}{{\sqrt
{1-\alpha_{11}^2-\alpha_{13}^2}}}\right],
\end{eqnarray}}
where different elements $\alpha_{ij}$ appearing in the rotation
matrix are defined in Eq.(21). It is easy to see that in the limit
$n \rightarrow \infty$, these angles precisely yield those of the
semiclassical model given in Eq.(14). This clearly shows that our
treatment of the quantized model is in conformity with the Bohr
correspondence principle and indicates the consistency of the
approach.
\par
The time-dependent probability amplitudes of the four levels are
given by \begin{equation}
 \left[ {\begin{array}{*{20}c}
    {C_1^{n + 2} (t)}   \\
   {C_2^{n + 1} (t)}   \\
    {C_3^n (t)}   \\
    {C_4^{n - 1} (t)}   \\
\end{array}} \right] = T_n ^{ - 1} \left[ {\begin{array}{*{20}c}
   {e^{ - i\lambda _{1q} t} } & 0 & 0 & 0  \\
   0 & {e^{ - i\lambda _{2q} t} } & 0 & 0  \\
   0 & 0 & {e^{ - i\lambda _{3q} t} } & 0  \\
   0 & 0 & 0 & {e^{ - i\lambda _{4q} t} }  \\
\end{array}} \right]T_n \left[ {\begin{array}{*{20}c}
    {C_1^{n + 2} (0)}   \\
   {C_2^{n + 1} (0)}   \\
    {C_3^n (0)}   \\
    {C_4^{n - 1} (0)}   \\
\end{array}} \right].
\end{equation}

In the next Section we proceed to analyze the probabilities of the
four levels for aforesaid initial conditions, namely,
\begin{eqnarray}
\nonumber
 {\begin{array}{*{20}l}
\textrm{Case-V} &: &C_1^{n+2} (0) = 1, &C_2^{n+1} (0)= 0,
&C_3^{n}(0)= 0, &C_4^{n-1} (0)= 0,\\
\textrm{Case-VI} &: &C_1^{n+2} (0) = 0, &C_2^{n+1} (0) = 1,
&C_3^{n}(0) = 0, &C_4^{n-1} (0) = 0,\\
\textrm{Case-VII} &: &C_1^{n+2} (0) = 0, &C_2^{n+1} (0) = 0,
&C_3^{n} (0) = 1, &C_4^{n-1} (0) = 0\\
\textrm{Case-VIII} &: &C_1^{n+2} (0) = 0, &C_2^{n+1} (0) = 0,
&C_3^{n} (0) = 0, &C_4^{n-1} (0) = 1, \end{array}}
\end{eqnarray}
respectively and then compare the results with those of the
semiclassical model.

\begin{center}
\large VI. Numerical results
\end{center}
\paragraph*{}
We are now in position to explore the physical content of our
treatment by comparing the probabilities of the semiclassical and
quantized cascade four-level system respectively. Fig.1a-d shows the
plots of the probabilities $|C_{1}^i(t)|^2$ (level-1, red line),
$|C_{2}^i(t)|^2$ (level-2, green line),  $|C_{3}^i(t)|^2$ (level-3,
blue line) and $|C_{4}^i(t)|^2$(level-4, black line) for the
semiclassical model corresponding to Case-I, II, III and IV
respectively. The comparison of Fig.1a (Fig.1b) and
\begin{figure} [h]
\begin{center}
\centerline{\epsffile{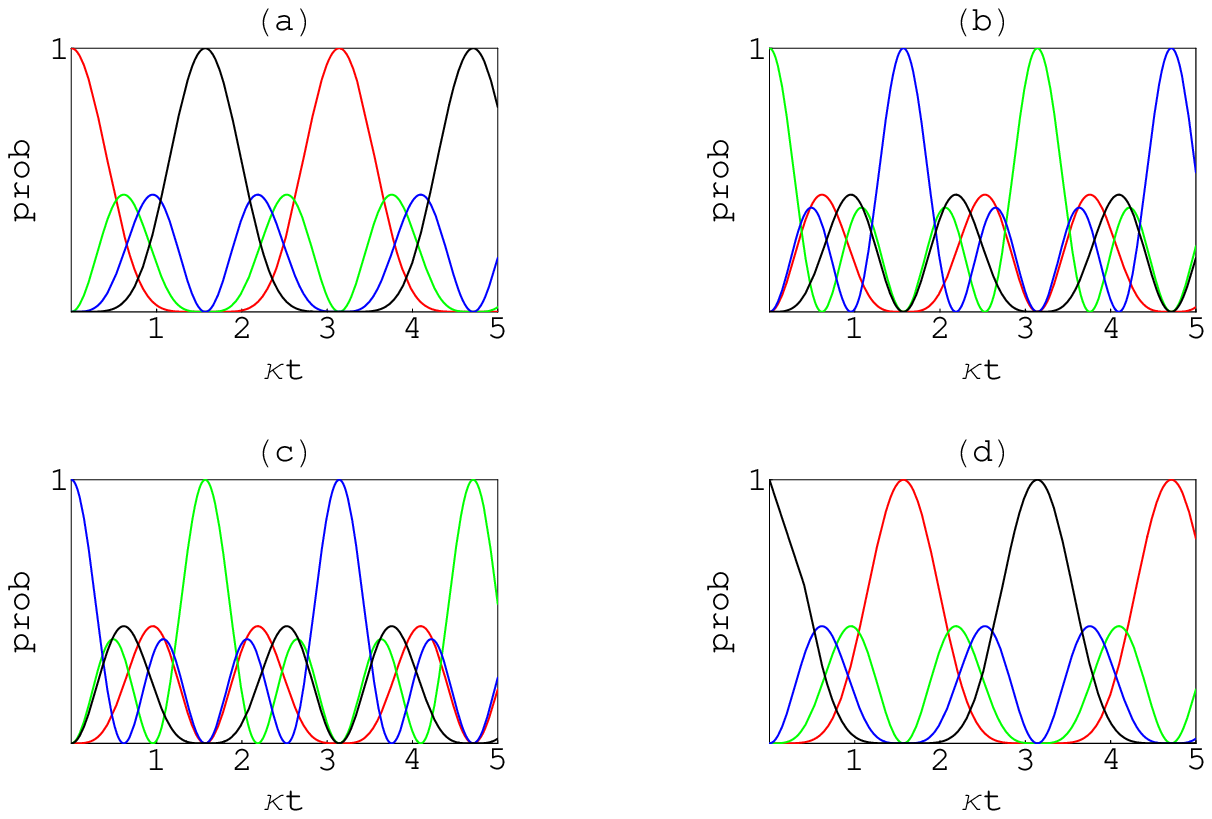}}
\end{center}
\noindent {\small {\bf [Fig.1]: For the semiclassical model, the
time evolution (scaled with $\kappa$) of the probabilities for
Case-I, II, III and IV are shown in Fig.(1a), (1b), (1c) and (1d),
respectively. The Rabi oscillation of Fig.(1a) and (1d) and that of
Fig.(1b) and (1c) are found to be similar. The probabilities of
level-1 (red line) and level-4 (black line) and those of level-2
(green line) and level-3 (blue line) are interchanged.}}
\end{figure}

Fig.1d (Fig.1c) shows that the pattern of the probability
oscillation of Case-I (Case-II) is similar to that of Case-IV
(Case-III) except the probabilities of level-1 and level-4 and also
that of level-2 and level-3 are interchanged. Thus a regular pattern
of the probability oscillation reveals the symmetric behavior of
Rabi oscillation for the semiclassical four-level cascade system.
\par
Following Ref.[44], in case of the quantized field, we consider the
time evolution of the probabilities for two distinct situations,
first, when the field is in a number state representation and then,
when the field is in the coherent state representation.
\par
For the number state representation, the Rabi oscillation for
Case-V, VI, VII and VIII of the quantized system are shown in
Fig.2a-d.
\begin{figure} [h]
\begin{center}
\centerline{\epsffile{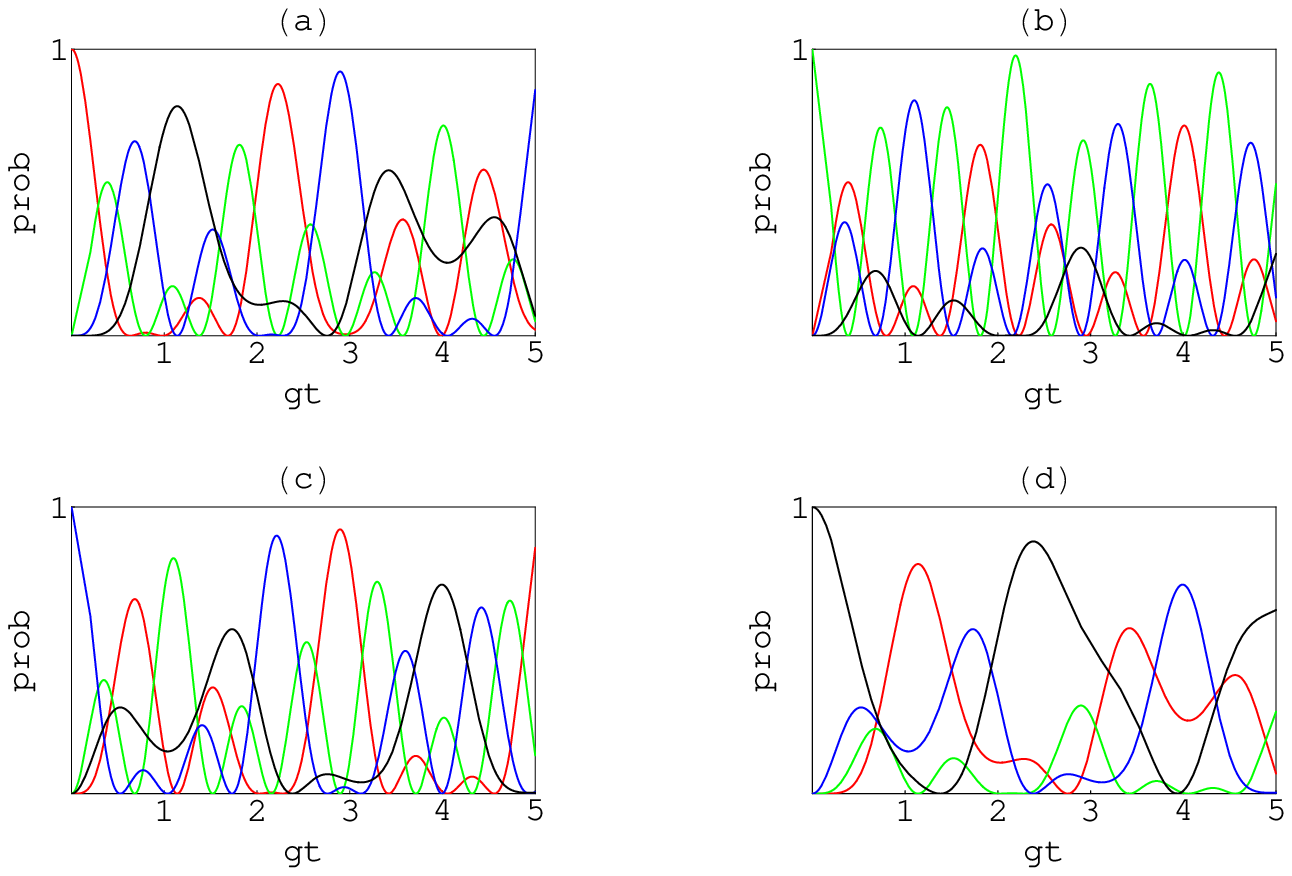}}
\end{center}
\noindent {\small {\bf [Fig.2]: The Rabi oscillation (scaled with
$g$) for Case-V, VI, VII and VIII with quantized cavity mode shows
the breaking of the aforesaid symmetry between
Case-I and Case-IV and between
 Case-II and Case-III, respectively.}}
\end{figure}
Here we note that for Case-V (Case-VI), the oscillation pattern of
the system is completely different from that of Case-VIII (Case-VI).
Thus the symmetry observed in the pattern of the population dynamics
of the semiclassical model between Case-I and Case-IV and also
between Case-II and Case-III no longer exists. In other word, for
the quantized field, in contrast to the semiclassical case, whether
the system initially stays in any one of the four levels, the
symmetry of the Rabi oscillation in all cases is completely spoilt.
As pointed out earlier, the disappearance of the symmetry is
essentially due to the vacuum fluctuation of the quantized cavity
mode which survives even at $n=0$. Recently we have reported similar
breaking pattern in the Rabi oscillation for the equidistant cascade
[44] and also for lambda and vee three-level systems [46]. Such
breaking is not observed in case of two-level Jaynes-Cumming model
and hence is essentially a nontrivial feature of  multi-level
systems when the number of levels exceeds two.
\par
Finally we consider the model interacting with the mono-chromatic
quantized field which is in the coherent state. The coherently
averaged probabilities of the system for level-1, level-2, level-3
and level-4 are given by
{\setlength\arraycolsep{2pt}
\begin{eqnarray}
\left\langle {P_1 (t)} \right\rangle   &=& \sum\limits_{n} {w_n }
\left| {C_1^{n + 2} (t)} \right|^2 \\
\left\langle {P_2 (t)} \right\rangle   &=& \sum\limits_{n} {w_n }
\left| {C_2^{n + 1} (t)} \right|^2\\
 \left\langle {P_3 (t)}\right\rangle   &=& \sum\limits_{n} {w_n } \left| {C_3^{n}
(t)} \right|^2 \\
\left\langle {P_4 (t)} \right\rangle   &=& \sum\limits_{n} {w_n }
\left| {C_4^{n - 1} (t)} \right|^2,
\end{eqnarray}}
respectively, where $w_n  = \exp [-\bar n]\frac{\bar{n}^n}{n!}$ be
the coherent distribution with $\bar n$ being the mean photon number
of the quantized field mode. Fig.3 and 4 display the numerical plots
of Eqs.(24-27) with $\bar n=48$ for Case-V, VI, VII and VIII,
respectively where the collapse and revival of the Rabi oscillation
is clearly evident. The collapse and revival for Case-V depicted in
Fig.3a-d is compared with that of Case-VIII shown in Fig.3e-h. We
note that Fig.3a, 3b, 3c and 3d are precisely identical to that of
Fig.3h, 3g, 3f and 3e respectively.

\begin{figure} [t]
\begin{center}
\centerline{\epsffile{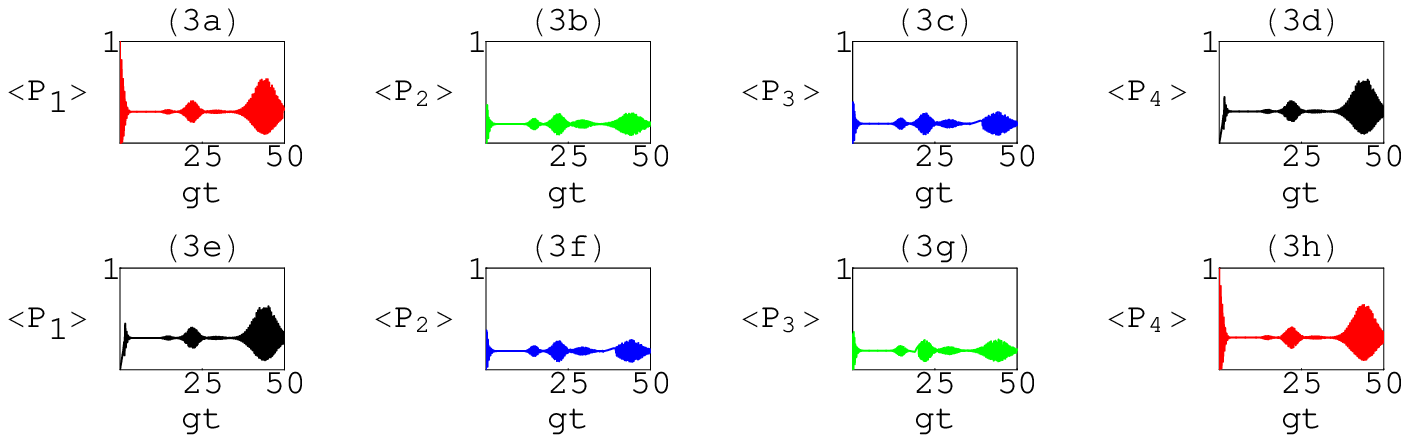}}
\end{center}
\noindent {\small {\bf [Fig.3]: Fig.3a-d and Fig.3e-h depict the
time-dependent collapse and revival phenomenon for Case-V and
Case-VIII respectively. We note that the oscillation pattern of for
level-1 (red), 2 (green), 3 (blue) and 4 (black) in Case-V is
similar to that of level-4, 3, 2 and 1 for Case-VIII,
respectively.}}
\end{figure}

\begin{figure} [h]
\begin{center}
\centerline{\epsffile{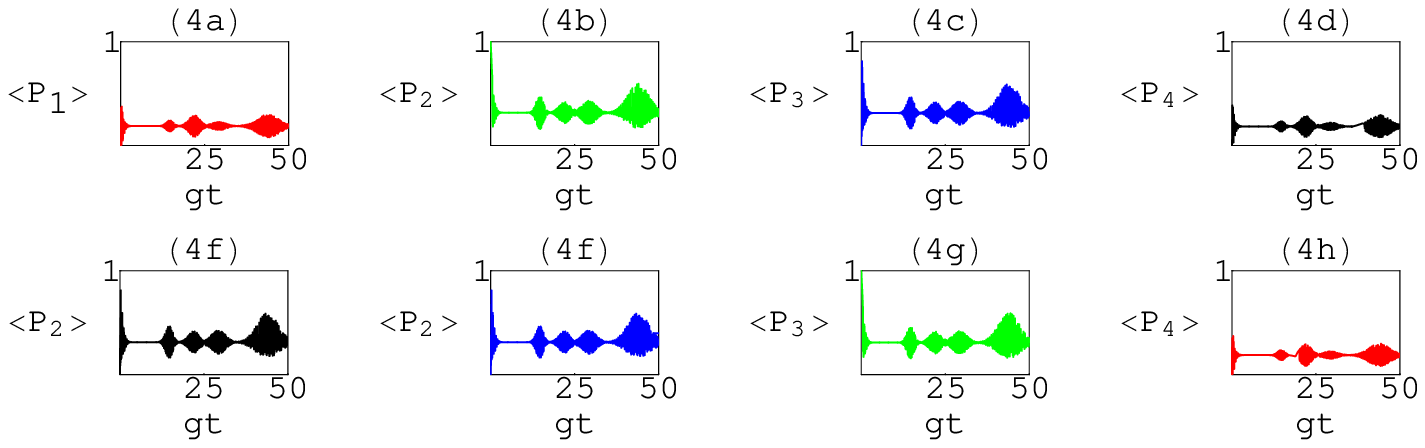}}
\end{center}
\noindent {\small {\bf [Fig.4]: Fig.4a-d and Fig.4e-h depict the
collapse and revival phenomenon for Case-VI and Case-VII,
respectively. Here we find that the oscillation pattern of level-1
(red), 2 (green), 3 (blue) and 4 (black) for Case-VI is similar to
that of level-4, 3, 2 and 1 for Case-VII, respectively.}}
\end{figure}
Similarly Fig.4 compares the collapse and revival of the system for
Case-VI with that of Case-VII, where, similar to the semiclassical
model, we note that Fig.4a, 4b, 4c and 4d are similar to that of
Fig.4h, 4g, 4f and 4e, respectively. Note that a distinct collapse
and revival pattern appears for a coherent cavity field only for a
large average photon number. This clearly recovers the symmetric
pattern exhibited by the semiclassical four-level system with the
classical field mode.
\begin{center}
\large V.Conclusion
\end{center}
\par
This paper examines the behaviour of the oscillation of probability
of a cascade four-level system taking the field to be either
classical or quantized. The Hamiltonian of the system is constructed
from the generators of the spin-$\frac{3}{2}$ representation of the
$SU(2)$ group and the probabilities of the four levels are computed
for different initial conditions using a generalized Euler angle
representation. We argue that the symmetry exhibited in the Rabi
oscillation with the classical field is completely destroyed due to
the quantum fluctuation  of the cavity mode. This symmetry is,
however, restored by taking the cavity mode as a coherent state with
large average photon number. It is interesting to look for the
effect of field quantization on the Rabi oscillation with other
configurations of the four-level system and to scrutinize its
non-trivial effect on the various coherent phenomena involving
multilevel systems.
\begin{center}
\large Acknowledgement
\end{center}
MRN and TKD thank University Grants Commission, New Delhi and SS
thanks Department of Science and Technology, New Delhi for partial
support. We thank Professor B Bagchi for bringing Ref.[45] to our
notice. SS is also thankful to S N Bose National Centre for Basic
Sciences, Kolkata for supporting his visit to the centre through the
Associateship Program. MRN and SS thank Dr A K Sen for his interest
in this problem.
\bibliographystyle{plain}

\begin{thebibliography}{5.8 in}
\bibitem[1] {Louisell1} W H Louisell, Quantum Statistical Properties of
Radiation, (Wiley, New York, 1973) pp 318
\bibitem[2] {Jaynes} E T Jaynes and F W Cummings, Proc. IEEE, {\bf 51}
(1963) 89
\bibitem[3] {Brewer} R G Brewer and E L Hahn, Phys Rev {\bf A11} (1975)
1641; P W Milloni and J H Eberly, J Chem Phys {\bf 68} (1978) 1602,
E M Belanov and I A Poluktov JETP {\bf 29} (1969) 758; D
Grischkowsky, M M T Loy and P F Liao, Phys Rev {\bf A12} (1975) 2514
and references therein
\bibitem[4]{Sobolewska} B Sobolewska, Opt Commun {\bf 19} (1976) 185, C
Cohen-Tannoudji and S Raynaud, J Phys {\bf B10} (1977) 365
\bibitem[5] {Whitley} R M Whitley and C R Stroud Jr, Phys Rev {\bf A14}
(1976) 1498
\bibitem[6]{Arimondo} E Arimondo, Coherent Population Trapping in Laser
Spectroscopy, Prog in Optics XXXV Edited by E Wolf (Elsevier
Science, Amsterdam, 1996) p257.
\bibitem[7] {Bowden} C M Bowden and C C Sung, Phys Rev {\bf A18} (1978)
1588; {\bf A20} 378(E)
\bibitem[8] {Mossberg} T W Mossberg, A Flusberg, R Kachru and S R
Hartman, Phys Rev Lett {\bf 39} (1984) 1523; T W Mossberg and S R
Hartman, Phys Rev {\bf 39} (1981) 1271
\bibitem[9] {Bergman} K Bergman, H Theuer and B W Shore, Rev Mod Phys
{\bf 70} (1998) 1003
\bibitem[10] {Cook} R J Cook and H J Kimble, Phys Rev Lett {\bf 54}
(1985) 1023; R J Cook, Phys Scr, {\bf T21} (1988) 49
\bibitem[11] {Misra} B Misra and E C G Sudarshan, J Math Phys {\bf 18}
(1977) 756-763; C B Chiu, E C G Sudarshan and B Misra, Phys Rev {\bf
D16} (1977) 520; R J Cook, Phys Scr, {\bf T21} (1988) 49
\bibitem[12] {Hioe} F T Hioe and J H Eberly, Phys Rev Lett {\bf47}
(1981) 838
\bibitem[13] {Dulcic} A Dulcic, Phys Rev {\bf A30} (1984) 2462
\bibitem[14] {Diels} J C Diels and S Besnainou, J Chem Phys {\bf 85}
(1986) 6347
\bibitem[15] {Smith} A V Smith, J Opt Soc Am {\bf B9} (1992) 1543
\bibitem[16] {Shore} B W Shore, The Theory of Coherent Atomic
Excitation (Wiley, New York, 1990)
\bibitem[17] {Allen} L Allen and J H Eberly, 1975 Optical Resonance and
Two-Level Atoms (Wiley, New York, 1975)
\bibitem[18] {Schmidt} H Schmidt and A Imamoglu, Opt Lett {\bf 21}
(1996) 1936
\bibitem[19]{Harris} S E Harris and Y Yamamoto, Phys Rev Lett {\bf 81},
(1998) 3611
\bibitem[20]{Lukin} M D Lukin, S F Yelin, M Fleischhauer and M O
Scully, Phys. Rev. {\bf A60}, (1999) 3225
\bibitem[21]{Korsunsky} E A Korsunsky and D V Kosachiov, Phys Rev {\bf
A60} (1999) 4996
\bibitem[22]{Yelin} S F Yelin and P R Hemmer, quant-ph/0012136.
\bibitem[23]{Paspalakis1} E Paspalakis and P L Knight, J Mod Opt
{\bf49} (2002) 87
\bibitem[24]{Paspalakis2} E Paspalakis and P L Knight, Phys Rev {\bf
A66} (2002) 025802
\bibitem[25]{Entin} D McGloin, D J Fulton and M H Dunn, Opt Commun
{\bf 190} (2001) 221
\bibitem[26]{KLHBW} E A Korsunsky, N Leinfellner, A Huss, S Baluschev
and L Windholz, Phys Rev {\bf A 59} (1999) 2302
\bibitem[27]{Yan} M Yan, E G Rickey and Y Zhu, Phys Rev {\bf A64}
(2001) 041801
\bibitem[28]{Chen} Y C Chen, Y A Liao, H Y Chiu, J J Su and I A Yu,
Phys Rev {\bf A64} (2001) 053806
\bibitem[29]{Badger} S D Badger, I G Hughes and C S Adams, J Phys {\bf
B34} (2001) L749
\bibitem[30]{Unanyan} R G Unanyan, B W Shore and K Bergmann, Phys Rev
{\bf A59} (1999) 2910
\bibitem[31] {Kis} Z Kis and F Renzoni, Phys Rev {\bf A65}
(2002) 032318
\bibitem[32] {Ham} B S Ham and P R Hemmer,  Phys Rev Lett {\bf 84}
(2000) 4080
\bibitem[33] {Agarwal} G S Agarwal and W Harshawardhan, Phys Rev Lett
{\bf 77} (1996) 1039
\bibitem[34] {Rau} A R P Rau, G Selvaraj and D Uskov, Phys Rev {\bf
 A71} (2005) 062316
\bibitem[35] {CS} R J Cook, B W Shore, Phys Rev {\bf A20} (1979) 539
\bibitem[36] {Bogolubov} N N Bogolubov Jr, F L Kien and A S Shumovsky,
Phys Lett {\bf 107A} (1985) 173
\bibitem[37] {Koz} M Kozierowski, J Phys B: At Mol Phys {\bf 19} (1986)
L535
\bibitem[38] {Buck} B Buck, C V Sukumar, J Phys {\bf A17} 877 (1984)
\bibitem[39] {Li} F Li, X Li, D L Lin and T F George, Phys Rev {\bf
A40} (1989) 5129 and references therein
\bibitem[40] {Fujii1} K Fujii, K Higashida, R Kato, T Y Suzuki and Y
 Wada,
quant-ph/0410003 v2
\bibitem[41] {Fujii2} K Fujii, K Higashida, R Kato, T Y Suzuki and Y
 Wada, quant-ph/0409068
\bibitem[42] {Liu} Z D Liu, S Y Zhu, X S Li, J Mod Opt {\bf 35} (1988)
 833
\bibitem[43] {Osman} K I Osmana and H A Ashi, Physica A 310 (2002) 165
\bibitem[44] {Nath1} Nath M R, Sen S and Gangopadhyay G, Pramana-J Phys
 {\bf 61}
(2003) 1089
\bibitem[45] {Bose} S K Bose and E A Pascos, Nucl Phys {\bf B169} 384
(1980); We have corrected the typos of this reference.
\bibitem[46] {Nath2} M R Nath, S Sen, G Gangopadhyay and A K Sen,
(Communicated)
\end{thebibliography}

\end{document}